# Decisive role of oxygen vacancy in ferroelectric vs. ferromagnetic Mn-doped $BaTiO_3$ thin films

Yao Shuai,[1] Shengqiang Zhou,[1,2] Danilo Bürger,[1] Helfried Reuther,[1] Ilona Skorupa,[1] Varun John,[1] Manfred Helm,[1] and Heidemarie Schmidt[1]

[1]*Institute of Ion Beam Physics and Materials Research, Helmholtz-Zentrum Dresden-Rossendorf, P. O. Box 510119, Dresden 01314, Germany*
[2]*State Key Laboratory of Nuclear Physics and Technology, School of Physics, Peking University, Beijing 100871, China*

**Abstract**: Single-phase perovskite 5 at.% Mn-doped and undoped polycrystalline $BaTiO_3$ thin films have been grown under different oxygen partial pressures by pulsed laser deposition on platinum-coated sapphire substrates. Ferroelectricity is only observed for the Mn-doped and undoped $BaTiO_3$ thin films grown under relatively high oxygen partial pressure. Compared to undoped $BaTiO_3$, Mn-doped $BaTiO_3$ reveals a low leakage current, increased dielectric loss, and a decreased dielectric constant. Ferromagnetism is seen on Mn-doped $BaTiO_3$ thin films prepared under low oxygen partial pressure and is attributed to the formation of bound magnetic polarons (BMPs). This BMP formation is enhanced by oxygen vacancies. The present work confirms a theoretical work from C. Ederer and N. Spaldin on ferroelectric perovskites [Nature Mat. 3, 849 (2004)] which shows that the existence of ferroelectricity is incompatible with the existence of a spontaneous magnetization in Mn-doped $BaTiO_3$ thin films.

## 1. INTRODUCTION

Much work has been carried out in the field of perovskite oxides due to their relatively simple structure and widespread applications including, for example, sensors, transducers, and memories. Among the perovskite oxides, $BaTiO_3$ has been one of the most intensively studied compounds for more than 60 years.[1-5] The structure of $BaTiO_3$ can be described as a set of $TO_6$ octahedra arranged in a cubic pattern, with the $Ba^{2+}$ ions located in the spaces between the octahedra and the $Ti^{4+}$ ions occupying the center of the octahedra.[6] In order to improve the electric properties of $BaTiO_3$, various dopants have been introduced into bulk or thin film $BaTiO_3$ to substitute the $Ba^{2+}$ or $Ti^{4+}$. Wu et al.[7] demonstrated that a proper concentration of Mn dopants in $BaTiO_3$-based ceramics can significantly increase the dielectric constant and decrease the dielectric loss, which was explained by the reaction: $Mn^{4+} + 2Ti^{3+} \rightarrow Mn^{2+} + 2Ti^{4+}$. Cole et al.[8] fabricated a 5 mol % Mg-doped $Ba_{1-x}Sr_xTiO_3$ multilayer structure and showed that Mg-doping reduces the dielectric loss of $Ba_{1-x}Sr_xTiO_3$ to 0.008 and yields a smoother surface compared to undoped $Ba_{1-x}Sr_xTiO_3$.

Meanwhile, an increasing interest has been devoted to so-called multiferroic materials which possess more than one type of ferroic order (ferromagnetic/antiferromagnetic, ferroelectric, ferroelastic, and ferrotoroidic).[9-13] $BaTiO_3$ has been proven both theoretically and experimentally to be multiferroic by doping or synthesis at the nanoscale.[14-20] Since ab-initio calculations[14] demonstrated the possibility to induce ferromagnetism in $BaTiO_3$ by doping with transition metal ions, much experimental work has been done by several groups. Lin et al.[16] prepared epitaxial Co-doped BaTiO$_3$ thin films on Nb-doped SrTiO$_3$ substrates by pulsed laser deposition and observed a ferromagnetic (FM) behavior at room temperature. The largest saturation magnetization of 187 emu/cm$^3$ was obtained by tuning the thickness of the Co-doped BaTiO$_3$ thin films. The origin of this ferromagnetism is still under discussion and needs more experimental investigations. Other researchers have reported on ferromagnetism in undoped BaTiO$_3$ ceramics and ascribed it solely to point defects such as oxygen vacancies.[20]

Although the doping in BaTiO$_3$ has been intensively investigated so far, most work either has focused on its electric or magnetic properties and did not consider their coexistence. In the present paper, 5 at.% Mn-doped BaTiO$_3$ polycrystalline thin films on Pt/sapphire have been investigated with an emphasis on the competition relationship between ferroelectric and ferromagnetic properties in dependence on Mn doping and deposition oxygen pressure.

## 2. EXPERIMENT

High-purity BaTiO$_3$ (4N) powder and a mixture of high purity BaTiO$_3$ (4N) and MnO$_2$ (5N) powder were pressed and finally sintered at 1300 °C for 12 hours to obtain the undoped and Mn-doped BaTiO$_3$ ceramic targets, respectively. Ca. 100 nm thick Pt bottom electrodes were sputtered on c-sapphire substrates outside the pulsed laser deposition (PLD) chamber. The BTMO and BTO thin films were grown by PLD using a KrF excimer laser ($\lambda$=248 nm) with a fixed repetition rate of 2 Hz and a deposition temperature of 650 °C. To investigate the dependence of the ambient-gas partial pressure on the electric and magnetic properties of the BTO and BTMO thin films, an oxygen partial pressure of 10 or 100 mTorr was applied. In the following the four samples presented in this paper (Tab. 1) are labeled BTO-10, BTO-100, BTMO-10, and BTMO-100, respectively. After deposition, the films were *in-situ* post-annealed for 30 minutes at 650°C and subsequently cooled down to room temperature with a cooling rate of 5 K/min under a constant oxygen partial pressure of 200 mbar. The thickness of the thin films was determined by step profilometry and strongly depends on the oxygen partial pressure and slightly on the Mn content. The PLD growth conditions and film thickness are summarized in Tab. 1.

The crystal structure of the films was analyzed by means of x-ray diffraction (XRD) using a Bruker D8 system with CuKα radiation. The valence state of Ti and oxygen was checked by x-ray photoelectron spectroscopy (XPS) measurement. For electric measurements circular Au top electrodes with a diameter of 0.2 mm were deposited by magnetron sputtering on top of the as-prepared BTO and BTMO thin films using a metal shadow mask to form a parallel-plate capacitor with BTO or BTMO as the dielectric medium. The resulting sample geometry is Au/BTO/Pt/sapphire and Au/BTMO/Pt/sapphire from top to bottom. The dielectric properties of the BTMO and BTO films were analyzed by an impedance analyzer (Agilent 4294A) in the frequency range from 100 Hz to 100 kHz under a driving voltage of 50 mV, the polarization-electric field loops of the films were observed using a Sawyer-Tower circuit[21] at 10 kHz. The magnetic properties of the films were investigated by a superconducting quantum interference device magnetometer (Quantum Design MPMS).

## 3. RESULTS AND DISCUSSION

### I. STRUCTURE INVESTIGATION

Figure 1 shows the XRD θ-2θ patterns of the BTMO and BTO thin films grown on Pt/c-sapphire substrates. Only diffraction peaks corresponding to the perovskite structure are observed, indicating that the prepared thin films have a single perovskite phase. The BTO-100 film is preferentially (111)-oriented [Fig. 1(a)], and the BTMO-100 film reveals a random orientation [Fig. 1(c)], while the BTO-10 film and BTMO-10 film are highly (001)-oriented [Fig. 1(b)(d)]. Both the doping and decrease of the deposition pressure have broadened the diffraction peaks significantly, indicating a deterioration of the thin film crystallinity. Although the BTMO films contain 5 at % Mn, only perovskite phases were detected in the BTMO-10 and BTMO-100 films [Fig. 1(c)(d)].

The lattice constants of the four samples are calculated from the XRD patterns and summarized in Tab. 2. It is clear that the Mn dopant induces a structure change from tetragonal to cubic. That transition is more obvious between BTO-100 and BTMO-100 which are grown at higher oxygen pressure. Note that the low oxygen growth pressure expands both the out-of-plane (c) and in-plane (a) lattice constants, but the in-plane lattice constant increases more significantly; therefore, a transition from tetragonal to cubic is also induced by decreasing the oxygen pressure during the growth.

### II. ELECTRIC PROPERTIES

In the following we discuss the influence of Mn doping and deposition pressure on the electric and magnetic properties in BaTiO$_3$. The I-V curves recorded on Au/BTO/Pt/c-sapphire and Au/BTMO/Pt/c-sapphire samples show symmetric characteristics, which rules out the interface-limited conduction. Mn doping together with high oxygen deposition pressure suppress the leakage current by two orders of magnitude, while low oxygen pressure leads to large leakage current even if Mn doping is introduced. [Fig. 2(a)] For better understanding the conduction mechanism of the thin films, the I-V curves are presented on a double logarithmic scale [Fig. 2(b)]. It is interesting that the BTO-100, BTMO-10 and BTMO-100 thin films show different conduction behavior. It has to be pointed out that the much higher leakage current of the BTO-10 thin film results from a shortening of all circular Au top contacts on this sample. The slope α of the log(I)-log(V) curve for the BTO-100 thin film is constantly close to 2, which implies that its leakage current is dominated by space-charge-limited conduction (SCLC).[22] While the log(I)-log(V) curve for the BTMO-10 and BTMO-100 thin films can be separated into two or three regions with different α of 2.8 and 8.4 for BTMO-10, 1.23, 1.84 and 6.13 for BTMO-100, respectively.

According to Rose's SCLC theory,[23] the BTO-100 thin film with a constant α ≈ 2 can be simply explained using a model of a trap-free insulator or an insulator with shallow traps which both give a relation between the current and voltage as I ∝ V$^2$. Also, the increase of α for the BTMO-10 and BTMO-100 thin films with increasing voltage involves one or two transition points which can also be interpreted by the SCLC mechanism.[23] It is well known that in perovskite oxides donor-like oxygen vacancies are present abundantly. Those oxygen vacancies provide a large concentration of electrons in the conduction band acting as free carriers and contribute to the leakage current. The Mn ions in BaTiO$_3$ have a fluctuating valence state between +2 and +4, which can compensate the oxygen vacancies to keep the sample electrical neutral.[24] Mn ions act as acceptors in the Mn-doped BaTiO$_3$ thin films and thus form positively charged trapping centers in the band gap. The trapping process can be described as follows: [25]

$$Mn_{Ti}^{\times} + e' \leftrightarrow Mn_{Ti}' \qquad (1)$$

$$Mn_{Ti}' + e' \leftrightarrow Mn_{Ti}'' \qquad (2)$$

The generated electrons in the films are preferentially trapped at Mn atomic sites, because $Mn_{Ti}^{\times}$ and $Mn_{Ti}'$ are more reducible than $Ti_{Ti}^{\times}$.[25] Therefore, the charge compensation for the presence of oxygen vacancies is probably attributed to the reduced valence state of Mn ions, but not the Ti ions, this speculation is confirmed by the XPS measurement which will be discussed later. In the low-voltage region, the Fermi level is fixed and the injected electrons from the electrode are trapped by the positive trapping centers formed by Mn ions as described by Eqs. (1-2). Only thermally activated carriers in the conduction band contribute to the current, resulting in an Ohmic conduction behavior. As the voltage increases, the Fermi level is shifted towards the conduction band, part of the trapping states are occupied by the injected electrons, while the remaining trapping centers are at thermal equilibrium with electrons in the conduction band. This leads to the first change of α to the value of 2, indicating SCLC conduction. When the voltage further increases, the second sharp increase of α occurs at the so-called trap-filled limit voltage ($V_{TFL}$) given by the following equation:[22]

$$V_{TFL} = qN_td^2/2\varepsilon_r\varepsilon_0, \qquad (3)$$

where $N_t$ is the trap density, d is the film thickness, $\varepsilon_r$ and $\varepsilon_0$ represent the static dielectric constant of the thin film and free space, respectively. The $V_{TFL}$ is 7.6 V and 10.4 V, and the static dielectric constant is measured to be 812 and 804 by the electrometer, for BTMO-100 and BTMO-10, respectively. Accordingly, the trap density $N_t$ is calculated to be $1.42\times10^{19}$ cm$^{-3}$ and $1.25\times10^{19}$ cm$^{-3}$, respectively, which is one order of magnitude higher than other groups reported for BaTiO$_3$-based systems.[26, 27] The high trap density indicates an effective substitution of Ti ions by Mn ions and a reduced tendency of dopant cluster formation as compared to other reports.[30] Beyond the $V_{TFL}$, all trapping states are occupied and the electrons are directly injected into the conduction band taking part in the conduction process and leading to a rapid increase of the current.[28] It should be noted that the Ohmic conduction behavior can not be observed on BTMO-10 and BTO-100, which is because the transition point from Ohmic conduction to SCLC conduction occurs at a very low voltage.

Based on the above discussion, the decrease of the leakage current in the BTMO-100 thin film is due to the formation of electron trapping centers caused by Mn acceptors in the band gap. Nevertheless, the BTMO-10 thin film shows leaky conduction even for high Mn doping concentration. This can be ascribed to the high concentration of free carriers offered by the large number of oxygen vacancies originating from the low oxygen pressure during growth.

Figure 3 shows the frequency dependence of the dielectric constant ($\varepsilon_r$) and the dielectric loss (tan$\delta$) of the BTO and BTMO thin films. The tan$\delta$ of the BTMO-10 thin film exhibits a noticeable drop within the low-frequency region, which indicates a large amount of oxygen vacancies in the thin film. On the other hand, it can be observed that the tan$\delta$ of BTMO-100 and BTO-100 thin films does not show large change over the whole frequency range. In addition, the BTMO-100 thin film displays higher tan$\delta$ than the BTO-100 thin film at low frequency. Chu *et al.*[29] have reported that Mn-doping can reduce the tan$\delta$ of BTMO thin films (the maximum concentration of Mn in their investigation was 1 at.%). On the contrary, in our case the tan$\delta$ of BTMO-100 thin film is slightly higher than the BTO-100 thin film grown under the same PLD growth condition. The higher tan$\delta$ of the BTMO-100 thin film can be accounted for by the existence of trapping centers in the band gap as compared to the BTO-100 thin film, which agrees with the presence of three different regions in the I-V curve. At low frequency, those trapping centers cause the increase of tan$\delta$.[24]

The $\varepsilon_r$ of the BTMO-100 and BTO-100 thin films is rather constant over the entire investigated frequency range. The $\varepsilon_r$ of the BTMO-100 thin films is smaller than that of the BTO-100 thin films, indicating that an excessively high concentration of Mn-doping decreases the $\varepsilon_r$ of BTMO thin films. This is consistent with the results of BTO ceramics reported by Wu *et al.*[3] The $\varepsilon_r$ of the BTMO-10 shows an obvious drop like the tanδ, also being due to oxygen vacancies which can respond to the small ac voltage only at lower frequency. By using a measurement frequency of 10 kHz (Fig. 3), the $\varepsilon_r$ and tanδ of the BTMO-10, BTMO-100 and BTO-100 thin films were found to be 537, 654, 882 and 0.044, 0.023, 0.016 respectively.

The P-E loops of the BTMO and BTO thin films measured by a Sawyer-Tower circuit at 10 kHz and an excitation ac signal of ±10 V are shown in Figure 4. The P-E loops indicate a typical ferroelectric (FE) behavior of both the BTMO-100 and BTO-100 thin films. The remnant polarization $P_r$ and coercive electric field $E_c$ are $2P_r$ = 6.0 μC/cm$^2$ and $2E_c$ = 64 kV/cm for the BTMO-100 thin film and $2P_r$ = 9.6 μC/cm$^2$ and $2E_c$ = 69 kV/cm for the BTO-100 thin film, respectively. The $2P_r$ of the BTMO-100 thin film is larger than that of other reported pure polycrystalline BTO thin films grown on different substrates,[30-32] and indicates a good quality of our samples. However, it is observed that the $P_r$ of BTMO-100 is smaller than that of BTO-100, indicating that introducing Mn dopants weakens the polarization of the thin film. That can be understood by the structure transition from ferroelectric, tetragonal BTO to paraelectric or weakly ferroelectric, cubic BTMO, which has been shown in Fig. 1 and Tab. 2. In a cubic lattice, the displacement of Ti$^{4+}$ is suppressed as compared to that in a tetragonal lattice, resulting in a smaller macroscopic polarization. Even larger remnant polarizations are expected at higher electric field for both BTMO-100 and BTO-100 thin films due to the fact that the leakage current density (Fig. 2) at the highest electric field amounts to only 1.5×10$^{-3}$ and 1.64×10$^{-3}$ A/cm$^2$, respectively. For the BTMO-10 thin film only a roundish PE curve has been obtained, in accordance with the large leakage current shown in the Fig. 2, which leads to the unsaturation of the PE curve.

### III. MAGNETIC PROPERTIES

The temperature-dependent magnetization-field (M-H) hysteresis loops of BTMO-10 are shown in Fig. 5(a). The BTMO-10 shows an obvious ferromagnetic (FM) behavior at room temperature. On the other hand, the BTMO-100 exhibits only paramagnetism at low temperature (Fig. 5(b)). In ref. 18, the room-temperature ferromagnetism of Mn-doped BaTiO$_3$ thin films was attributed to bound magnetic polarons (BMPs)[33] formed by localized electrons and Mn$^{4+}$, which has been widely used to explain the ferromagnetic behavior in diluted magnetic semiconductors. Xu *et al.*[17] observed ferromagnetism in Fe-doped BaTiO$_3$ ceramics and accounted for their results in terms of a model of carrier-mediated Zener-type Ruderman-Kittel-Kasuya-Yosida interaction.[34] Lee *et al.*[15] obtained ferromagnetic 5 at.% Mn-implanted single crystalline BaTiO$_3$ at 10 K, but they were not able to determine whether the ferromagnetism was due to bound magnetic polarons or carrier-induced magnetism in the Zener-type mechanism. In particular, Tong *et al.*[23] synthesized non-stoichiometric Mn-doped BaTiO$_3$ nanoparticles with Ti-deficiency and observed ferromagnetism at low temperature, and they explained it by means of a ferromagnetic exchange effect between holes trapped at Ti vacancies and magnetic Mn ions. Their result agrees with the conclusion of Nakayama *et al.*[14] that hole-doped Ba(Ti$_{1-x}$Mn$_x$)O$_3$ should be ferromagnetic. Besides the transition metal ion-doped BaTiO$_3$ thin films and bulk materials, Mangalam *et al.*[20] reported room temperature ferromagnetism in pure BaTiO$_3$ nanoparticles and ascribed it to oxygen vacancies which were identified by employing positron annihilation spectroscopy. The complexity of the different ferromagnetic coupling mechanisms of doped or virgin BaTiO$_3$ is therefore due to the different preparation processes.

We observed a distinct difference in the M-H loops of BTMO-10 and BTMO-100 which reveals that the ferromagnetism strongly depends on the oxygen vacancy concentration in the samples. Oxygen vacancies are preferentially formed with oxygen underpressure during the PLD growth process. The observed ferromagnetism in BTMO-10, in contrast to the paramagnetism in BTMO-100, indicates that oxygen vacancies are essential for introducing ferromagnetism into BaTiO$_3$ thin films. Moreover, BTMO-10 and BTO-10 are believed to contain the comparable concentration of oxygen vacancies, because they were grown at the same oxygen pressure, but BTO-10 does not show any magnetism (inset of Fig. 5(a)). Therefore, based on the above two evidences, it is reasonable to conclude that both doping with Mn and the presence of oxygen vacancies play important roles in generating ferromagnetism in BaTiO$_3$. Lin *et al.*[16] reported significantly enhanced magnetism in their Co-doped BaTiO$_3$ as the thickness decreased, and attributed this to the possibility of more carriers being present in thinner films. However, they did not discuss the I-V of their samples. Our M-H curves along with the I-V curve clearly show that not only the doping with transition metal ions, but also the oxygen vacancy concentration is a decisive factor to introduce room-temperature ferromagnetism in BaTiO$_3$ thin films.

To elucidate the origin of the ferromagnetism in BTMO-10, XPS measurements were performed on BTMO-10 and BTMO-100. It is expected that Ti ions at 3+ and lower valence state may have magnetic ordering. For example, it is observed that $Ti^{3+}$ in $TiO_2$ contributes to the ferromagnetism.[35] However, as shown in the insets of Fig. 6(a) and (b), the Ti has the same valence state of 4+ in both samples which is indicated by the $Ti_{2p3/2}$ peak appearing at 458.32 eV and 458.45 eV, respectively.[36] From that, we conclude that Ti in Mn-doped $BaTiO_3$ does not contribute to the observed ferromagnetism.

However, a large difference is observed in the O 1s spectra [Fig. 6(a) and (b)]. The peak at 529.6 eV (peak1) corresponds to the lattice oxygen, and they have comparable intensity. While the peak located at 532.4 eV (peak2) is significantly different. The increased intensity of the peak2 in BTMO-10 appears to be due to a larger concentration of oxygen vacancies in BTMO-10. It has to be pointed out that there is a debate on the interpretation of the O 1s XPS spectra probed on oxides. Note that the appearance of the peak2 can also be due to surface contamination.[37,38] However, BTMO-10 and BTMO-100 have a very similar fabrication process and the only difference is the oxygen under pressure during deposition. Therefore, no large difference in the surface contamination is expected. In the I-V and C-V measurement, it is clear that more oxygen vacancies are present in BTMO-10. Thus it is reasonable that the stronger peak2 in BTMO-10 is due to the increased oxygen vacancy concentration.[39,40]

Although the valence state of Mn ions is not possible to be measured due to the detection limitation of the instrument, based on the O 1s spectra and the same valence state of Ti in both samples, the charge compensation for the presence of oxygen vacancies is undertaken by the valence reduction of Mn ions, in other words, Mn ions may have multiple valence states, as has been discussed in the I-V measurement. The reduction of the Mn valence state is essentially due to the electron trapping by oxygen vacancy, which has also been proved by fitting the I-V curve. The Mn ions and trapped electrons form BMPs.[40,41] The large concentration of oxygen vacancies in BTMO-10 leads to a sufficient large number of BMPs which overlap with each other, thus creating a long range ferromagnetic order in BTMO-10. On the other hand, the absence of ferromagnetism in BTMO-100 is ascribed to non-coupled Mn ions because of the reduced oxygen vacancies. The similar observation has also been reported in Co/Mn-doped $BaTiO_3$ or $TiO_2$, in which an enhanced ferromagnetism was obtained by increasing the concentration of oxygen vacancies or of other defects.[18,43-45]

### IV. MODELING AND DISCUSSION

Note that even for the BTMO-10 thin film with a high oxygen vacancy concentration, not all Mn dopant ions are ferromagnetically coupled. The measured magnetic moment is attributed to isolated paramagnetic and to ferromagnetically coupled Mn dopant ions which can be observed in Fig. 5(a). The paramagnetic component dramatically decreases with increasing temperature from 5 K and disappears at 150 K.

In order to determine the valence state and concentration of isolated paramagnetic Mn ions, the paramagnetic components of BTMO-10 and BTMO-100 at 5 K were fit by the Brillouin function $M=NgJ\mu_B B_J(y)$, where $y=gJ\mu_B H/k_B T$, $B_J(y)=\frac{2J+1}{2J}\coth(\frac{2J+1}{2J}y)-\frac{1}{2J}\coth(\frac{y}{2J})$. N is the concentration of isolated Mn ions, which is the fitting parameter together with L. S was chosen to be 5/2, 2, or 3/2 in dependence on the valence state of the Mn ion which was varied during the fitting. For fitting the contribution from isolated Mn ions to the paramagnetism, the ferromagnetic component in BTMO-10 and BTMO-100 was subtracted using the M-H loop measured at 300 K, in which only a ferromagnetic component exists. The fitting results are shown in Fig. 7(a) and (b), the fitting parameters are summarized in Tab. 3. It is seen that for BTMO-10, the reasonable fitting parameters are only obtained for $Mn^{2+}$ where the fitted orbital angular momentum L amounts to 0.1. For $Mn^{4+}$ ions the fitted concentration N value is much larger than the expected value in 5 at.% Mn-doped $BaTiO_3$. On the other hand, in BTMO-100 the more reasonable fitting was obtained when we assumed all the Mn ions are $Mn^{3+}$ or $Mn^{4+}$. The fitting result shows a clear trend that more $Mn^{2+}$ ions are present in BTMO-10, while in BTMO-100 $Mn^{3+}$ or $Mn^{4+}$ ions are more favorable. In 5 at.% Mn-doped $BaTiO_3$, the Mn ion density is expected to be $7.6\times10^{20}$ cm$^{-3}$. The N value obtained from the fitting is one order of magnitude smaller than the expected value. The smaller N in Tab. 2 is attributed to two possible reasons: (1) the presence of Mn clusters although no impurity phase has been observed in XRD pattern. (2) antiferromagnetically coupling of some Mn ions.

The fitting result clarifies that in BTMO-10 more $Mn^{2+}$ ions are present due to the fact that more oxygen vacancies exist in this sample as compared to BTMO-100, which has been proved in the I-V, C-V, and P-E measurement. Therefore, our results support that the ferromagnetism in BTMO-10 is strongly correlated to the BMPs formation which is caused by the presence of oxygen vacancy.

Even though ferromagnetism has been obtained in BTMO-10, the small $M_S$ value compared to other typical ferromagnetic materials should be noticed, and at the same time its ferroelectricity is deteriorated by the large leakage current caused by the oxygen vacancies. Ederer and Spaldin[46] have explained the difficulty to realize multiferroic perovskites and they clarified the exclusivity between ferromagnetic and ferroelectric order in $BaTiO_3$. To stabilize the off-centering of Ti ions which is essential for ferroelectricity, an energy-lowering covalent bond has to be formed, which requires empty *d* orbitals of the transition metal ions. However, the substitution of Ti ions by Mn ions destroys the stabilization of the off-centering displacement, which weakens the polarization as shown in the PE loops of BTMO-100 due to a structure transition from tetragonal to cubic. On the other hand, a partially occupied *d* shell of transition metal ions is needed for magnetization. Moreover, the ferromagnetic coupling between Mn ions happens, if a large concentration of oxygen vacancies is present, which increases the leakage current and thus deteriorates PE further (BTMO-10).

## 4. CONCLUSIONS

In conclusion, polycrystalline 5 at.% Mn-doped $BaTiO_3$ and undoped $BaTiO_3$ thin films were prepared on Pt/sapphire substrates at 100 mTorr and 10 mTorr oxygen pressures by pulsed laser deposition. The thin film shows a transition from tegragonal to cubic structure caused by Mn dopants and the low oxygen pressure during the growth. The electric properties strongly depend on the oxygen vacancy concentration controlled by the ambient pressure during PLD and are also significantly influenced by doping with Mn. The leakage current has been decreased by the substitution of Ti with Mn ions which form trapping centers in the $BaTiO_3$ band gap. The dielectric constant decreased and dielectric loss increased as compared to undoped $BaTiO_3$ thin film. The Mn-doped $BaTiO_3$ films with a large amount of oxygen vacancies show room-temperature ferromagnetic behavior which can be explained by the formation of BMPs. On the contrary, neither the Mn-doped $BaTiO_3$ thin film with low oxygen vacancy concentration nor the pure $BaTiO_3$ thin film with high oxygen vacancy concentration exhibit ferromagnetism. Although the doping with Mn has decreased the leakage current, which is very important for device application, the ferroelectricity has been impaired. Besides, sole Mn doping is not enough for introducing room-temperature ferromagnetism, but oxygen vacancies are also necessary. Therefore, ferroelectricity and ferromagnetism are excluding each other mutually in Mn-doped $BaTiO_3$

## ACKNOWLEDGMENTS

Y. S. would like to thank the China Scholarship Council (grant number: 2009607011). S. Y, D. B. and H. S. thank the financial support from the Bundesministerium für Bildung und Forschung (BMBF grant number: 13N10144).

**FIGURE CAPTIONS**

Fig. 1    XRD patterns of (a) BTO-100 (b) BTO-10 (c) BTMO-100, and (d) BTMO-10 thin films on Pt/sapphire substrates. XRD reflexes are labeled by the symbols o, ●, and ◆ and represent Au, Pt, and c-sapphire, respectively.

Fig. 2    Leakage current density of the BTO-100, BTMO-100, BTO-10, and BTMO-10 thin films.

Fig. 3    Frequency dependence of the dielectric constant ($\varepsilon_r$) and dielectric loss (tan$\delta$) of the BTO-100, BTMO-100, and BTMO-10 thin films.

Fig. 4   Polarization-electric field hysteresis loops of the BTO-100, BTMO-100 and BTMO-10 thin films.

Fig. 5   Magnetization-field hysteresis loops of the BTMO-10 and BTMO-100. The inset of Fig. 5(a) shows M-H loop of BTO-10 at 5 K.

Fig. 6    O1s XPS spectrum of (a) BTMO-10, (b) BTMO-100. The insets show the Ti$_{2p}$ spectrum of BTMO-10 and BTMO-100, respectively.

Fig. 7    Fitting the paramagnetic component for (a) BTMO-10 and (b) BTMO-100 using the Brillouin function.

Table.1 Sample identification and corresponding PLD growth conditions and film thickness in units of nm.

| Sample | Temperature (°C) | Pressure (mTorr) | Thickness (nm) |
| --- | --- | --- | --- |
| BTMO-10 | 650 | 10 | 272 |
| BTMO-100 | 650 | 10 | 219 |
| BTO-10 | 650 | 100 | 281 |
| BTO-100 | 650 | 100 | 224 |

Table. 2 Lattice constants (in Å) of the BTO and BTMO thin films calculated from the XRD patterns.

| sample | c (out-of-plane) | a (in-plane) | c/a |
| --- | --- | --- | --- |
| BTMO-10 | 4.049 | 4.046 | 1.0007 |
| BTMO-100 | 4.014 | 4.012 | 1.0005 |
| BTO-10 | 4.047 | 4.033 | 1.0035 |
| BTO-100 | 4.028 | 3.982 | 1.0116 |

Table. 3 Expected and fitting parameters. N: Mn ion density. L: orbital momentum. S: spin momentum. N: ion concentration.

| | Mn$^{2+}$ | | | Mn$^{3+}$ | | | Mn$^{4+}$ | | |
|---|---|---|---|---|---|---|---|---|---|
| | L | S | N (cm$^{-3}$) | L | S | N (cm$^{-3}$) | L | S | N (cm$^{-3}$) |
| Expected | 0 | 5/2 | 7.6×10$^{20}$ | 1.5 | 2 | 7.6×10$^{20}$ | 3 | 3/2 | 7.6×10$^{20}$ |
| BTMO-10 | 0.1 | 5/2 | 8.2×10$^{19}$ | -0.3 | 2 | 7.1×10$^{19}$ | 0.4 | 3/2 | 3.1×10$^{21}$ |
| BTMO-100 | -1.6 | 5/2 | 7.2×10$^{19}$ | 3.4 | 2 | 6.6×10$^{19}$ | 3.3 | 3/2 | 4.7×10$^{19}$ |

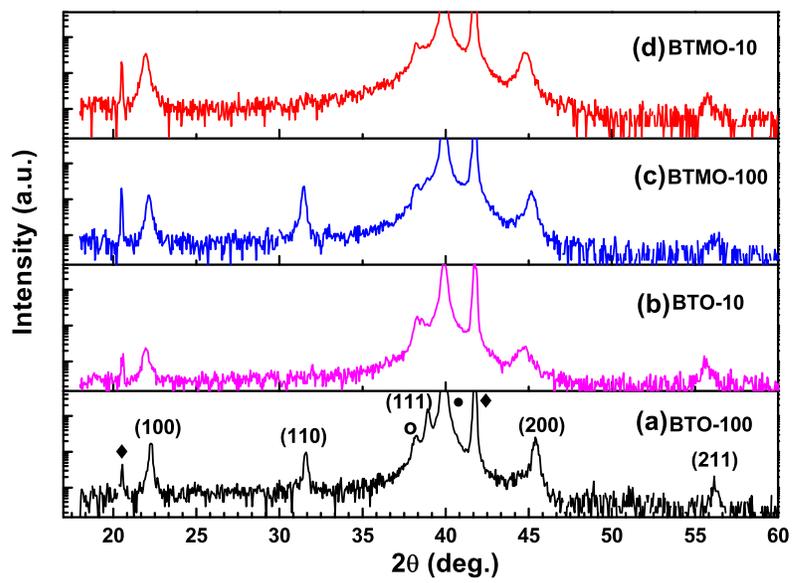

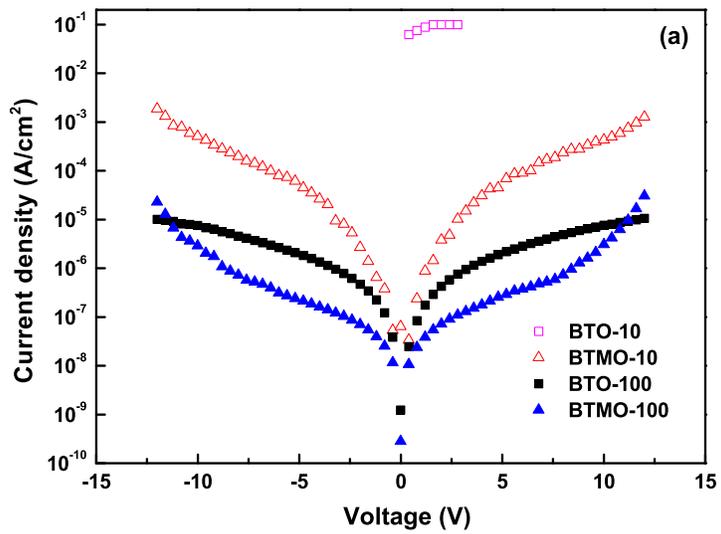

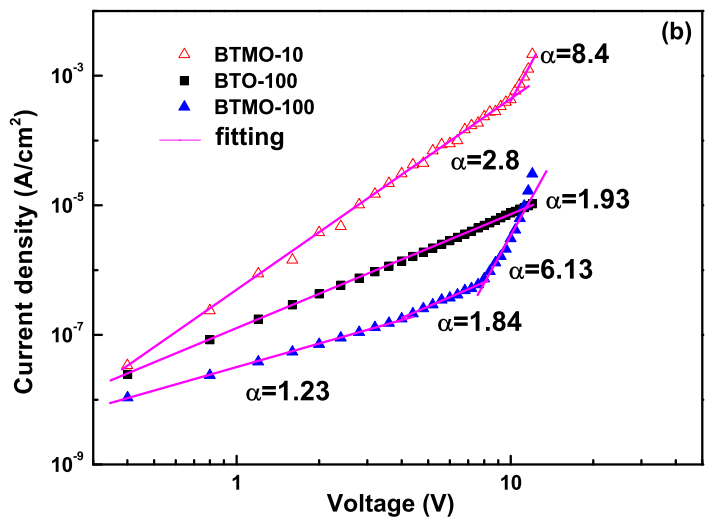

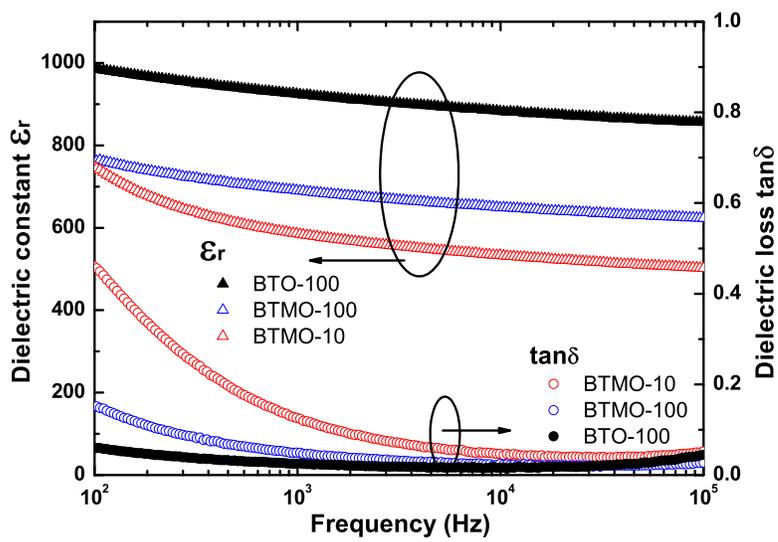

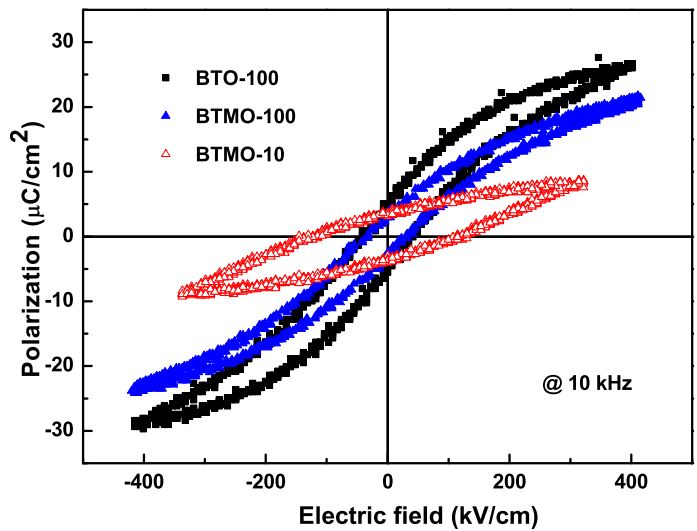

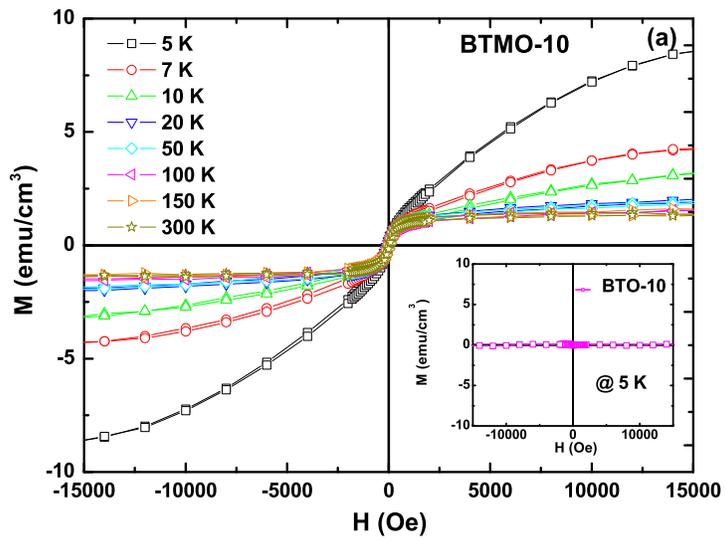

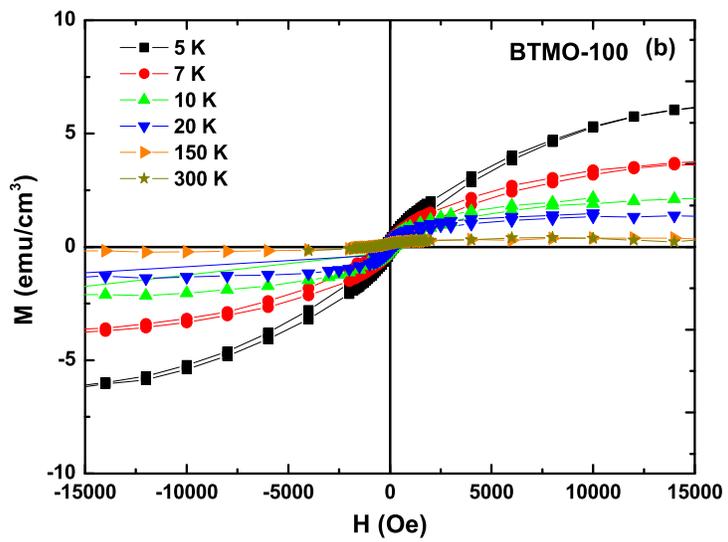

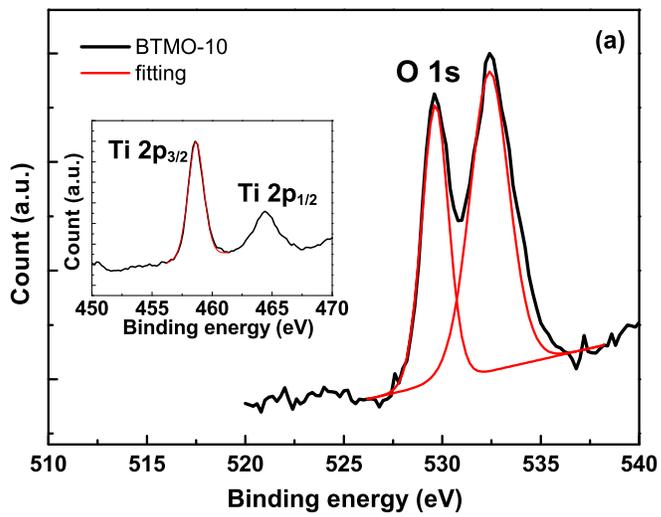

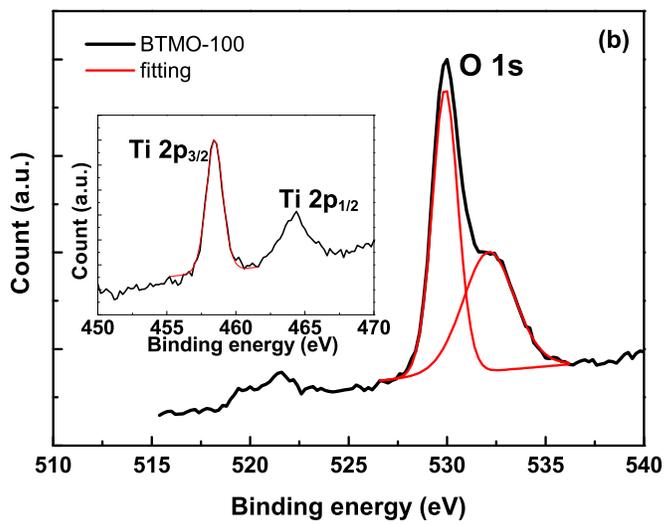

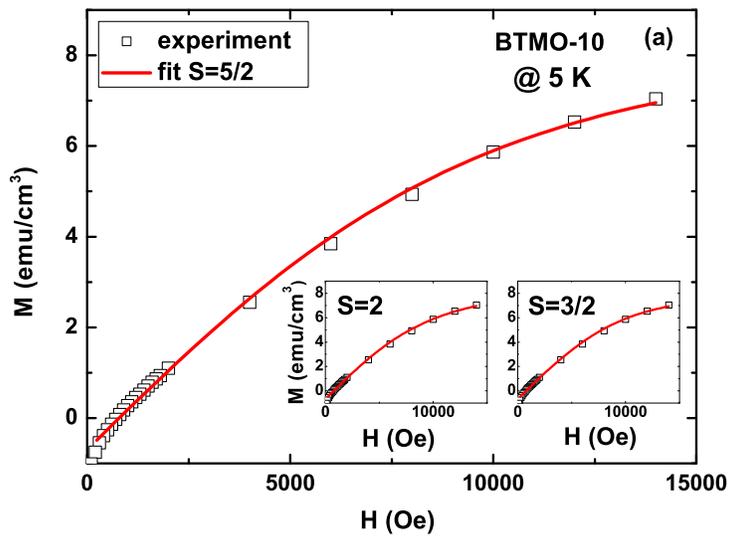

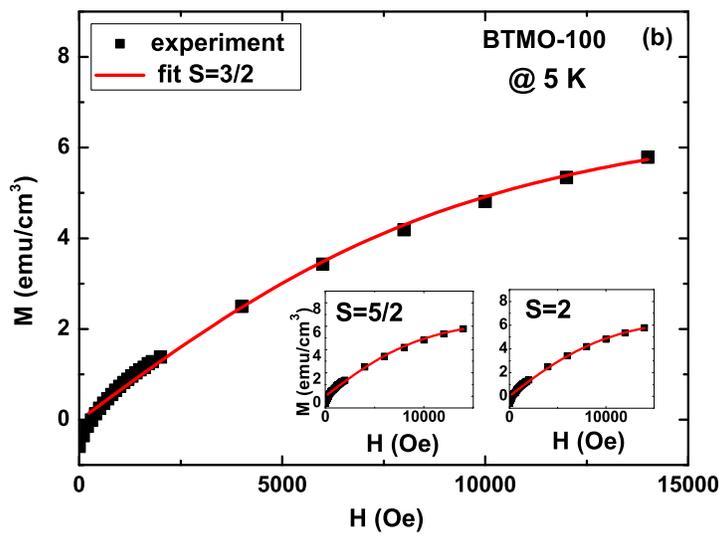